\documentstyle[mnras_cite]{mn}
\def\spose#1{\hbox to 0pt{#1\hss}}
\def\simlt{\mathrel{\spose{\lower 3pt\hbox{$\mathchar"218$}}
     \raise 2.0pt\hbox{$\mathchar"13C$}}}
\def\simgt{\mathrel{\spose{\lower 3pt\hbox{$\mathchar"218$}}
     \raise 2.0pt\hbox{$\mathchar"13E$}}}

\def\hmpc{\;h^{-1}{\rm Mpc}}

\def\kms{{\rm km\;s}^{-1}}

\def\etal{{\it et~al. }}

\newcommand{\lqt}{\stackrel{<}{_{\sim}}}
\newcommand{\gqt}{\stackrel{>}{_{\sim}}}
\def\hmpcrev{\;{\rm h}\;{\rm Mpc}^{-1}}

\begin{document}
\title[The Power Spectrum of Rich Clusters of Galaxies on Large
Spatial Scales] {The Power Spectrum of Rich Clusters of Galaxies on
Large Spatial Scales} 

\author[H. Tadros, G. Efstathiou and G. B. Dalton]{Helen
Tadros$^{1}$ , 
George Efstathiou$^{2}$, and Gavin Dalton$^{2}$.
\\
$^1$Astronomy Centre, University of Sussex,
Falmer, Brighton BN1 9QH, U.K.\\
$^2$Department of Physics, University of Oxford,  
Keble Road, Oxford, OX1 3RH, UK.
}

\maketitle

\begin{abstract}
We present an analysis of the redshift-space power spectrum,
$P(k)$, of rich clusters of galaxies based on an automated cluster
catalogue selected from the APM Galaxy Survey.  We find that $P(k)$ 
can be approximated by a power law,
$P(k) \propto k^{n}$, with $n \approx -1.6$ 
over the wavenumber range $0.04\hmpcrev < k <
0.1\hmpcrev$.  Over this range of wavenumbers, the APM cluster
power spectrum has the same shape as the power spectra measured
for optical and IRAS galaxies. This is consistent with a simple
linear bias model in which 
different tracers have the same power spectrum as that of the 
mass distribution but shifted in amplitude by a constant
biasing factor. On larger scales, the power spectrum of APM
clusters  flattens  and appears
to turn over on a scale $k \sim 0.03\hmpcrev$. 
We compare the
power spectra estimated from simulated APM cluster catalogues to 
those estimated directly from cubical N-body simulation volumes and find that
the APM cluster survey should give reliable estimates of the true
power spectrum at wavenumbers $k \simgt 0.02\hmpcrev$. These
results suggest that the observed turn-over in the power spectrum
may be a real feature of the cluster distribution and that we have
detected the transition to a near scale-invariant power spectrum 
implied by observations of anisotropies in the cosmic
microwave background radiation. The scale of the  turn-over
in the cluster power spectrum is in good agreement with the
scale of the turn-over observed in the power spectrum of APM 
galaxies.
\end{abstract}

\section{Introduction}
\label{ssintro}

Most modern theories of structure formation, including those based on
inflation or topological defects in the early Universe, predict a near
scale-invariant spectrum of density perturbations on large spatial
scales. On small spatial scales, $\lambda \simlt 10 \hmpc$, the
observed fluctuations in the galaxy distribution differ very
substantially from a scale-invariant form. For example, 
the power spectra estimated from
infra-red and optically selected redshift surveys are reasonably well
approximated by $P(k) \propto k^{-1.5}$ at wavenumbers $k \simgt 0.1
\hmpcrev$ and there is no convincing evidence for a turn-over to a
scale-invariant form, $P(k) \propto k$, at smaller wavenumbers. However, 
such a turn-over must exist since measurements of the
microwave background fluctuations on angular scales $\simgt 1^{\circ}$
suggest a near scale-invariant spectrum of perturbations on spatial
scales $\simgt 100 \hmpc$ (see the reviews by \pcite{SSW95} and
\pcite{Bond96}). 

A convincing detection of a turn-over in the power spectrum of
density irregularities is therefore of fundamental
significance and would establish a direct
link between fluctuations observed in the microwave background
radiation and those observed in the density distribution. 
Furthermore, in theories such as the Cold Dark Matter (CDM)
model, a peak in the power spectrum is predicted on the scale of the
Hubble radius at the time that matter and radiation have equal
density, $\lambda_{equ} \sim 10 (\Omega h^{2})^{-1} {\rm Mpc}$ (see
e.g. \pcite{Efst90}). An observation of a peak in the power spectrum
of galaxy clustering can therefore be used to constrain the parameter
$\Gamma = \Omega h$ that is fundamental to CDM models.

There is some tentative evidence for a turn-over at a wavenumber $k
\sim 0.025$ in the three-dimensional power spectrum inferred from the
two-dimensional clustering of galaxies measured from the APM galaxy
survey (\pcite{BEI}, \pcite{BEII}, see for example Figure 11 of
\pcite{BEII}). \scite{GB97} have carefully investigated the
significance of this observed turn-over by repeating, using
simulations of the APM galaxy survey, the procedure of inverting the
two-dimensional clustering statistics to obtain the three-dimensional
power spectrum . They conclude that the turn-over, which occurs
between $0.02 < k < 0.06  \hmpcrev$, can be reproduced
by the inversion process. However, at these small
wavenumbers,  the systematic errors in the construction 
of the APM Galaxy Survey are
difficult to quantify and could perhaps be greater than the
random errors (see \pcite{APMlong} for a detailed discussion
of systematic errors in the APM survey).  It is
therefore difficult to assign a meaningful statistical significance to this
result. 

The detection of a peak in the power spectrum of the
galaxy distribution is one of the main scientific goals of the
Anglo-Australian 2dF redshift survey (see e.g. \pcite{Ef96}) and the
Sloan Digital Sky survey (\pcite{GW95}) which aim to measure redshifts
of $\sim 10^6$ galaxies. However, in this paper
we show that redshift surveys of much smaller numbers ($\sim 10^3$)
of rich clusters of galaxies can provide accurate measurements of the
power spectrum on large spatial scales. Relatively little work has
been done on the three-dimensional power spectrum of rich clusters of
galaxies (see \pcite{GE92}, \pcite{PW92}, \pcite{EGST}). Most recent
work has focused on measurements of the two-point spatial correlation
function of rich clusters of galaxies, $\xi(r)$, and whether
systematic errors in the cluster catalogues affect the amplitude of
$\xi(r)$ on scales $\simlt 20 \hmpc$. The closest investigation to the
one presented here is that of Peacock and Nicholson (1991), who
describe a power spectrum analysis of a redshift survey of $310$ radio
galaxies. Their results are described in further detail in Section 4.

In this paper we analyse the redshift survey of $364$ APM clusters
described by \scite{DCESMD94}.  In previous papers (\pcite{DEMS92}, 
\pcite{DCESMD94}), we have applied a number of tests to show that the
APM cluster catalogue is free of the projection and selection biases
known to affect clustering in the Abell cluster catalogues (\pcite{S88},
\pcite{EDSM92}). As we show here, the large volume surveyed by
the APM cluster survey ($V \simgt 3 \times 10^7 h^{-3} {\rm Mpc}^3$)
renders it suitable for an investigation of clustering on very 
large scales.

This paper is laid out as follows. In Section 2 we summarize the
techniques used in the power spectrum analysis. We apply these
techniques to the  APM cluster survey and investigate the
sensitivity of our results to the selection function, weighting
function and background cosmological model. In Section 3 we construct
simulated APM cluster catalogues from large N-body simulations to
quantify any systematic errors and biases in our power spectrum 
estimator. From
this analysis we can assess the significance of the observed peak in
power spectrum of APM clusters. Our conclusions are presented in
Section 4.

\section{Estimation of the power spectrum}
\label{ssps}

\begin{figure}
\centering
\begin{picture}(130,130)
\includegraphics{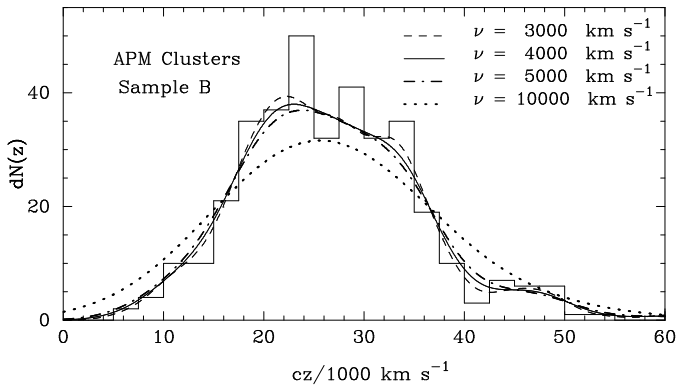}
\end{picture}
\caption{\label{nz} Histogram of the redshift distribution of the
$364$ clusters in Sample B of the APM cluster redshift
survey. The lines shows the result of smoothing the histogram
with a Gaussian of width $\nu$ for four values in the range 
$3000$ -- $10000 \;\kms$ as indicated in the figure.}
\end{figure}

The analysis described here is similar to the
power spectrum analysis of the Stromlo-APM galaxy redshift survey
described by \scite{tad96} (hereafter TE96). We follow
their notation unless otherwise stated. 
We apply the power spectrum analysis to the redshift survey of APM
clusters described by \scite{DCESMD94}. We use the largest statistically
uniform sample of APM clusters, Sample B of \scite{DCESMD94} containing
$364$ clusters over the southern APM area ($21^{h}\lqt RA \lqt5^{h},
-72.5^{\circ}\lqt dec \lqt-17.5^{\circ}$). The number density of clusters
in this sample is $\sim 3.4\times 10^{-5} (\hmpc)^{-3}$. The redshift
distribution of the sample is shown in Figure~\ref{nz}. The median
redshift of the sample is $z_{med} = 0.09$ and the cluster distribution
extends to a redshift  $z  \sim 0.2$.

\subsection{Method}
We treat the APM cluster redshift survey in the same way as a
flux limited galaxy survey, and use the methods of \scite{FKP94}
(hereafter FKP), as implemented by TE96, to
account for the radially varying selection function. To define the
selection function of the survey we smooth the velocity distribution
shown in Figure~\ref{nz} with a Gaussian of width $\nu\; \kms$. The 
fiducial value of $\nu$ is $4000 \;\kms$.

\begin{figure*}
\centering
\begin{picture}(300,280)
\includegraphics{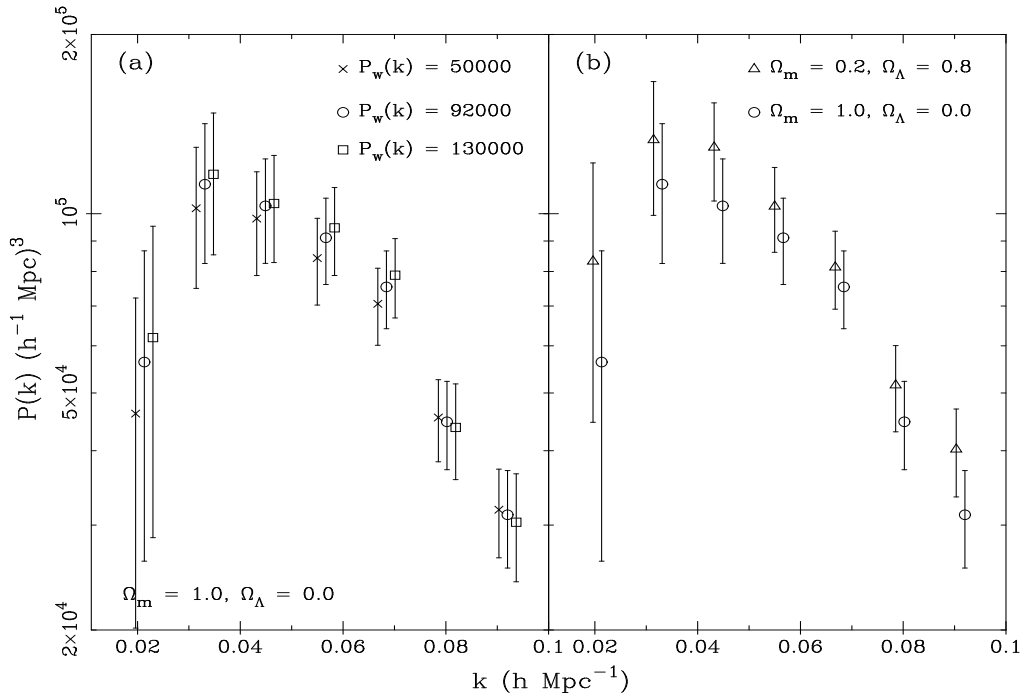}
\end{picture}
\caption[junk]{\label{clusps1} Three-dimensional power spectra of the
APM clusters redshift survey. Panel (a) shows the effect of using
different values of $P_{w}(k)$ in the weighting function of equation
(3). Three values are used: $P_{w}(k) = 50000, 92000$ and
$130000\left(\hmpc\right)^{3}$ as shown by the crosses, circles, and
squares respectively. The estimates in Panel (a) assume a spatially
flat geometry with $\Omega_\Lambda = 0$. Panel (b) shows the effect of
varying the assumed cosmology. The power spectrum is shown for two
spatially flat models: $\Omega_m = 1.0, \Omega_{\Lambda} = 0.0$
(circles), and $\Omega_m = 0.2, \Omega_{\Lambda} = 0.8$ (triangles).
Both spectra in the right hand panel have $P_{w}(k) = 92000
\left(\hmpc\right)^{3}$ in the weighting function. For clarity, sets
of points have been offset from each other by one seventh of the bin
spacing in both panels.}
\end{figure*}

The estimation of the power spectrum may be summarized as follows (for
further details the reader is referred to TE96).
We compute a weighted density field
\begin{equation}
F\left({\bf{r}}\right) =
\frac{w\left(r\right)\left[n_{c}\left({\bf{r}}\right) - \alpha
n_{s}\left({\bf{r}}\right)\right]}{\left[\int\overline{n}^{2}\left({\bf{r}}\right)w^{2}\left(r\right)
d^{3}r\right]^{\frac{1}{2}}} \;,
\end{equation}
where the subscript $c$ denotes the cluster density in the real
catalogue and $s$ denotes the density field for a random catalogue 
with same angular and radial selection functions as the cluster survey.
In equation (1),
$\overline{n}\left({\bf{r}}\right)$ is the expected mean density of
clusters in a catalogue with the same angular and radial selection
functions as the data. The radial selection function is derived from
the fits to the redshift distribution plotted in Figure~\ref{nz}. The
function $\overline{n}\left({\bf{r}}\right)$ can be separated into
the mean galaxy density $\overline{n}\left(r\right)$ as a function
of radial distance $r$, multiplied by the angular mask of the 
catalogue
({\it i.e.} the distribution of the APM galaxies on the sky
defined by the survey boundary and excluding regions around bright
stars, globular clusters {\it etc}.). The factor
 $\alpha$ is the ratio of the space densities
in the real catalogue to that in the random catalogue. In the analysis
presented here we use several thousand times as many points in the
random catalogue as there are clusters in the real catalogue, and
compute $\alpha$ from the ratio of the sums
\begin{equation} 
\sum_i {1 \over (1 + 4 \pi \overline n(r_{i})J_3)} \; \;,
\end{equation}
where we have set $4 \pi J_3 = 40000 (h^{-1} {\rm Mpc})^3$ and the sums
run over all clusters and random points (see Efstathiou 1996, Section 5.3
for details). The specific choice for
$J_3$ is based on the power spectra of CDM models multiplied by a
factor $\sim 4$ to account for the bias of clusters with respect to
the underlying mass distribution. However, the results presented below
are insensitive to the exact value of $4 \pi J_3$ used in the analysis.

Each cluster in the analysis is weighted according to the minimum
variance weight function (FKP)
\begin{equation}
w\left(r\right) = \frac{1}{1 +
\overline{n}\left(r\right)P_{w}\left(k\right)}\; \; ,
\end{equation}
which depends on the value of $P_{w}(k)$ at each wavenumber $k$.  We
have set $P_{w}\left(k\right)$ in equation (3) equal to a constant
value and investigated how the estimates of the power spectrum change
for three values of $P_{w}\left(k\right)$:  $50000$,  $92000$ and
$130,000 \left(\hmpc\right)^{3}$, which span 
the range of the observed power spectrum 
over a wide range of wavenumbers. The power
spectrum is derived by Fourier transforming equation (1), removing
shot noise (equation 2.1.9 of FKP) and averaging the power-spectrum
estimates over shells in $k$--space of volume $V_k$. The power spectrum
estimates will be correlated over a range in $k$--space given
approximately by $\delta k \sim D^{-1} $ where D is the characteristic
depth of the survey. To obtain roughly independent points, all
power spectra presented in this paper (except those calculated from
the full periodic box of the N-body simulations described in
Section 3) have been binned over a range in wavenumber $\sim D^{-1}$.

To perform the power spectrum analysis one must define a background
cosmological model. Firstly, a cosmology must be assumed in the conversion
of cluster redshifts to proper distances, and secondly a cosmological
volume element must be used to obtain the radial number density from the
solid line in Figure~\ref{nz}.  Since the median redshift of clusters in
the APM survey $z_{med} = 0.09$, the assumed cosmology may affect the
results. Consequently we have carried out the analysis for two assumed
cosmologies. Firstly a spatially flat universe with a matter density
parameter $\Omega_m = 1.0$, and secondly a spatially flat universe with
$\Omega_m = 0.2$, and a cosmological constant contributing most of the
closure density, $\Omega_{\Lambda} = 0.8$. The results for an open
universe with $\Omega_m = 0.2$ and $\Omega_\Lambda =0$ are intermediate
between those for the two spatially flat models.

\subsection{Computation of Errors}
Error bars on $P(k)$ are computed using equation (2.4.6) of
\scite{FKP94}. Their expression was derived assuming that the density
field of the tracer objects followed a Gaussian distribution. This will
not be true on small scales where the density fluctuations in the cluster
distribution are greater than unity, even if the primordial fluctuations
were strictly Gaussian. Thus, in Section 3.2 we test the accuracy of the 
error estimates using mock cluster catalogues generated from large $N$-body
simulations.

\subsection{Results for $P(k)$}
In Figure~\ref{clusps1} we show the power spectrum for the $364$
clusters in the APM cluster redshift survey. The cluster sample was
truncated at a proper distance of $r_{max} = 600 \hmpc$ and the
selection function for the sample was derived by smoothing the cluster
redshift distribution with a Gaussian of width $4000 \kms$ as
illustrated in Figure 1.

Figure~2(a) shows the cluster power
spectra calculated assuming values of $P_{w}(k) = 50000, 92000$ and
$130,000\left(\hmpc\right)^{3}$ in the weighting function (equation 3)
and distances computed assuming a spatially flat universe with 
zero cosmological constant. Figure~2(b)
shows the effect of 
changing the underlying cosmological model whilst keeping the
weighting parameter fixed at $P_{w}(k) = 92,000
\left(\hmpc\right)^{3}$.  The circles show the power spectrum
derived assuming a spatially flat cosmology with $\Omega_m = 1.0$,
$\Omega_\Lambda=0$ and the
triangles  show the power spectrum calculated assuming $\Omega_m =
0.2, \Omega_{\Lambda} = 0.8$.  This comparison shows that
uncertainties in the background cosmological model can affect the
amplitude of the cluster power spectrum by a factor of up to $\sim
1.3$, but that the shape of the power spectrum is not significantly
affected. Figure 2 shows that the power spectrum of APM clusters is
well approximated by a power law $P(k) \propto k^n$
with index $n = -1.6 \pm 0.3$ over
the wavenumber range $0.04 < k < 0.1 \hmpcrev$. The power spectrum
flattens at smaller wavenumbers and appears to turn-over a wavenumber
$k \sim 0.03 \hmpcrev$.

\subsection{Sensitivity of $P(k)$ to weighting and selection
function}

Figure~\ref{clusps1} shows that the power spectrum of APM clusters is
not very sensitive to the value of $P_{w}(k)$ assumed in the weighting
function of equation (3). Changing the weighting factor
$P_w(k)$ by relatively large amounts leads to changes in the 
power spectrum points that are very much smaller than the error bars.

The power spectrum estimates are more sensitive to the to the value of
$\nu$, the velocity smoothing parameter that is used to generate a
smooth selection function from the cluster redshift distribution (see
Figure 1).  Figure 3(a) shows power spectra calculated using three
values for this smoothing, $3000 \kms$ and $5000 \kms$ (spanning the
fiducial value of $4000 \kms$ used in the estimates of Figure 2) and a
much larger value, $\nu = 10000 \kms$. For these estimates we use a
constant weighting factor of $P_{w}(k) = 92000
\left(\hmpc\right)^{3}$.  Although the power spectra are insensitive
to the smoothing velocity for small changes around the fiducial value
of $4000 \kms$, if we smooth with $\nu = 10,000 \kms$ the amplitude of
the power spectrum at wavenumbers $k \lqt 0.035$ rises sufficiently to
eliminate the turnover in $P(k)$. For the largest smoothing, the
selection function is a poor fit to observed redshift distribution,
and particularly to the small number of high redshift clusters ($z >
0.13$, see Figure~\ref{nz}).  We find that the sensitivity of the
power spectrum estimates to the selection function smoothing can be
reduced substantially by truncating the sample at a proper distance of
$r_{max} = 400 \hmpc$, thus eliminating the high redshift tail of the
cluster $n(z)$ distribution. This is illustrated in Figure 3(b) which
shows $P(k)$ estimates for the same three velocity smoothings as in
Figure 3(a) but where we have truncated the cluster sample at $r_{max}
= 400 \hmpc$. In this figure the three estimates of $P(k)$ are
compatible to well within the $1\sigma$ error bars. The effect of the
velocity smoothing is to produce a slight change in the amplitude of
$P(k)$ and all three estimates show a turnover at $k \sim 0.03
\hmpcrev$.

\begin{figure}
\centering
\begin{picture}(250,230)
\includegraphics{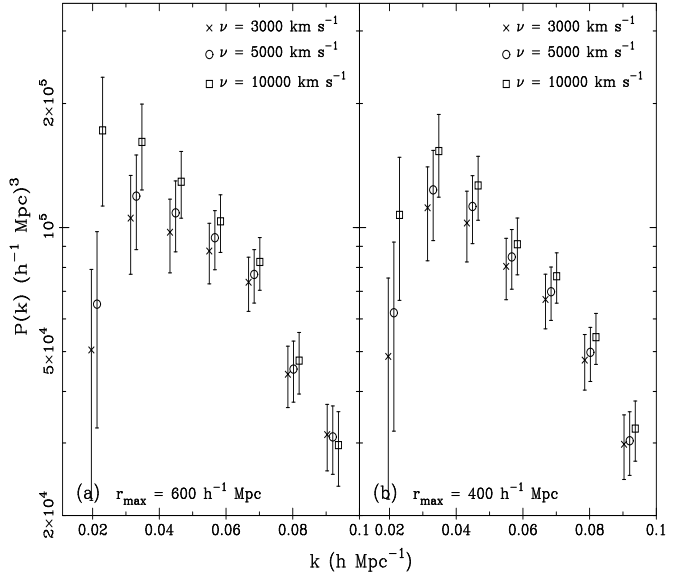}
\end{picture}
\caption{\label{clusps2} Figure 3(a) shows the effect on the 
power spectrum estimates of changing the 
value of $\nu$ used to smooth the cluster redshift 
histogram (see Figure1). Estimates of $P(k)$ are shown for three values 
of $\nu$: $\nu = 3000\; \kms$ (crosses), $\nu = 5000\; \kms$ (circles)
and $\nu = 10000 \;\kms$ (squares). The power spectrum
becomes sensitive to the value of $\nu$ for wavenumbers $k \lqt 0.035
\hmpcrev$. Figure 3(b) is identical to Figure 3(a) except that we
have limited the cluster sample to 
clusters with  proper distances of less than $400 \hmpc$. For all of these
estimates, we have assumed a constant weighting factor of 
$P_{w}(k) = 92000\left(\hmpc\right)^{3}$ and a spatially flat
background cosmology with $\Lambda=0$.  For clarity, sets
of points have been offset from each other by one seventh of the bin
spacing in both panels.}
\end{figure}

Unless otherwise stated, in the rest of this paper we will use
the APM cluster power spectrum computed under following
assumptions:
\begin{enumerate}
\item $P(k) = 92000 \left(\hmpc\right)^{3}$ in the weighting
function of equation (3).
\item A smoothing velocity $\nu = 4000 \;\kms$ to define the selection
function.
\item $r_{max} = 400 \hmpc$.
\item A spatially flat background cosmology with zero cosmological 
constant.
\end{enumerate}
To summarize, this section shows that the shape of the
estimated  power spectrum is insensitive to these assumptions
to within the $1\sigma$ error bars. However, plausible variations
of these assumptions can lead to a change in the 
overall amplitude of $P(k)$  by up to a factor of 
$1.5$.

\section{Comparison with simulated APM cluster surveys}
In this section we investigate the reliability of the power spectrum
estimator described in Section 2.1. We measure $P(k)$ from mock APM
cluster surveys extracted from large N-body simulations to
determine the minimum wavenumber at which the true power spectrum 
can be recovered accurately. From this investigation, we can assess the
significance of the turn-over seen in the power spectrum of APM
clusters at a wavenumber of $k \sim 0.03 \hmpcrev$ (Figures 2 and 3).

\subsection{Construction of simulated surveys}

We have constructed mock APM cluster redshift surveys from $N$-body
simulations of two cold dark matter (CDM) models.  The methods for
creating mock catalogue from the numerical simulations are described
in more detail by Croft and Efstathiou (1994).  The simulations consist of two
ensembles of 9 simulations each containing $160^{3}$ particles within
a periodic computational box of length $\ell_b = 600 \hmpc$.  They are
similar to the simulations described in detail by Croft and Efstathiou
(1994)  but
employ more particles within a larger computation box. (The simulations of
Croft and Efstathiou use $100^{3}$ particles within a box of size $\ell_b = 
300 \hmpc$).
The simulations were run with the particle-particle-particle-mesh
(P$^{3}$M) code described by
\scite{EDFW85} and model gravitational clustering in a 
CDM dominated universe with scale invariant initial density
fluctuations.  The two ensembles are as follows: the standard CDM
model (\pcite{DEFW85}), {\it i.e.} a spatially flat universe with
$\Omega_{0}=1$ and $h=0.5$ (the SCDM ensemble);  a spatially flat low density
CDM universe with $\Omega_{0}=0.2$, $h=1.0$, and a cosmological
constant contribution $\Omega_{\Lambda}= \Lambda /(3H_{0}^{2})
=\left(1-\Omega_{0}\right)=0.8$ (the LCDM ensemble). 
The initial power spectra of the models are generated from 
the fitting function
\begin{equation}
P\left(k\right)\propto
\frac{k}{\left[1+\left(ak+ \left(bk\right)^{\frac{3}{2}}+
\left(ck\right)^{2} \right)^{\nu}\right]^{\frac{2}{\nu}}},
\end{equation}
where $\nu=1.13$,
$a= (6.4/\Gamma) \hmpc$, $b=(3.0/\Gamma)\hmpc$ and
$c=(1.7/\Gamma) \hmpc$. Equation (4) is a good
approximation to the linear power spectrum of scale-invariant CDM
models with low baryon density, $\Omega_{b}\ll \Omega_{0}$
(\pcite{BE84}). The parameter $\Gamma$ in equation
(4) is equal to $\Omega_0h$. Thus $\Gamma = 0.5$ for the
SCDM ensemble and $\Gamma = 0.2$ for the LCDM ensemble.

\begin{figure*}
\centering
\begin{picture}(300,250)
\includegraphics{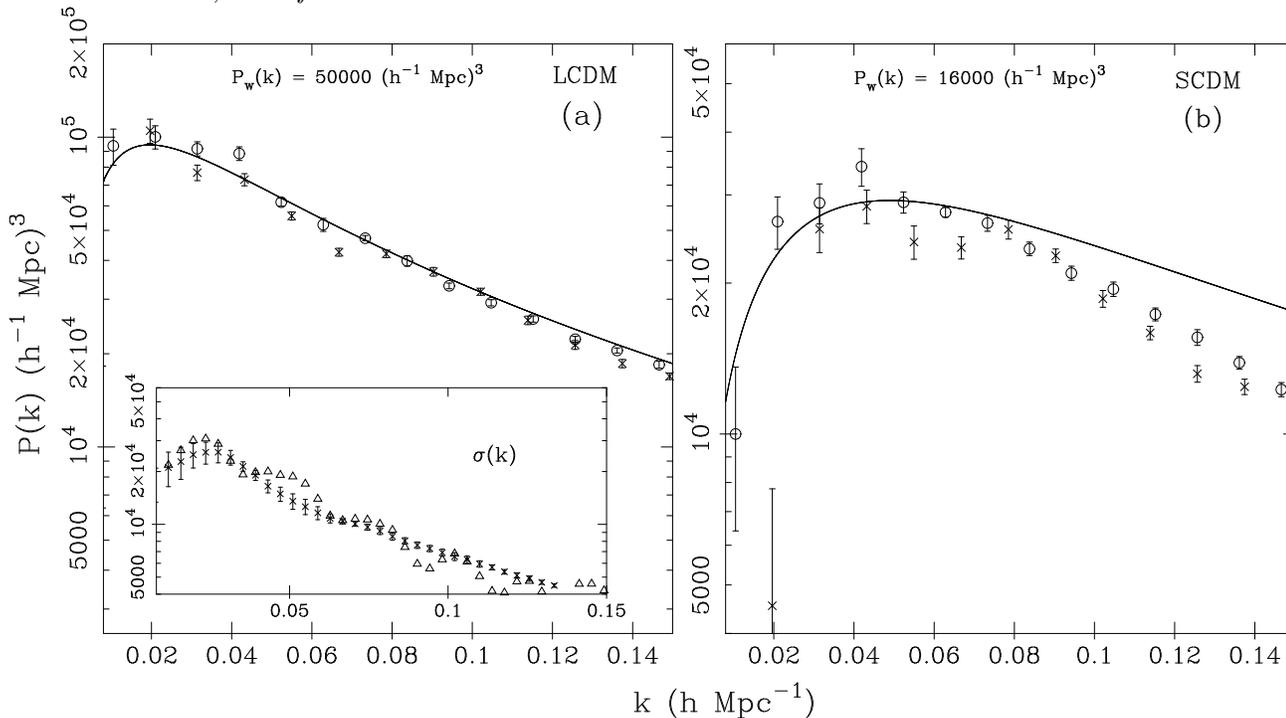}
\end{picture}
\caption[junk]{\label{plotpsrectest} The circles plotted in each of
the two panels show the redshift-space power spectrum of clusters in
the LCDM model (left hand panel) and the SCDM model (right hand panel)
computed from the full periodic volume of the N-body simulations.  The
crosses show the power spectra estimated from the average over $27$
realizations of the APM cluster redshift survey constructed from the
simulations. The error bars in all cases show the error on the mean
value of $P(k)$. The power spectrum estimated from mock APM cluster
catalogues is in good agreement with the power spectrum computed from
the full simulation volumes on scales $k \ge 0.02 \hmpcrev$ for the
LCDM model, and $k \ge 0.03 \hmpcrev$ for the SCDM model. The solid
lines show the linear theory spectra for the two models normalized
arbitrarily to match the data points at wavenumbers $k \sim 0.05
\hmpcrev$. The inset in the left hand panel shows the performance of
the error estimator (equation 2.4.6 of FKP). The triangles in the
inset plot show the error bar on the mean $P(k)$ obtained from the
$27$ realizations of the mock APM catalogues. The crosses show the
error, $\sigma(k)$, derived by using equation (2.4.6) of FKP. The
error bar plotted on each of the crosses shows the size of the error
on the mean error, obtained from the spread in the measured
$\sigma(k)$ over the $27$ realizations of the APM cluster catalogue.}
\end{figure*}

The final output times of the models are chosen to approximately match
the microwave background anisotropies measured in the first year COBE
maps (\pcite{COBE}) ignoring any contribution from gravitational
waves. Thus the {\it rms} mass fluctuations in spheres of radius $8
\hmpc$ are $\sigma_8 = 1$.  The clustering statistics of clusters of
galaxies are a very weak function of time, so the simulation output
time used is not a critical consideration (see e.g. \pcite{CE94a}).

Candidate cluster centres are selected by using 
a friends-of-friends algorithm
to link particles with separation less than $0.1$ times the mean
inter-particle separation. These group centres are then used as the
centres of spheres of size $r_{c} = 0.5\hmpc$ within which the mass of
the cluster is calculated.  The centre of mass of the cluster was
found and any cluster which was less than $r_{c}$ from the centre of
another, richer cluster was deleted.  The procedure was then repeated,
and a final  list of clusters was generated and 
ordered by mass. A lower mass cut 
off was applied to generate a cluster sample of a chosen mean number density.

The mock surveys extracted from the cluster distribution are designed
to sample the same volume as the APM cluster survey using the same sky
mask as the real data ({\it i.e.} the identical area of the sky and
excluded areas such as holes around bright stars and globular
clusters). The lower mass cut-off of clusters identified in the N-body
simulations was adjusted to match the space-density of clusters in the
real APM cluster redshift survey. The mock catalogues were limited at
a proper distance of $600 \hmpc$; however we have checked that the
results and conclusions of this section are unchanged if the mock
catalogues are limited at the smaller distance of $400 \hmpc$. The
smoothed APM cluster selection function ($\nu = 4000 \kms$) was
imposed on the cluster catalogues and (since the peculiar velocities
of the clusters in the N-body simulations are known) we shifted the
clusters to their redshift space positions to create mock APM
catalogues in redshift space.

\subsection{$P(k)$ from simulated surveys}

Figure~\ref{plotpsrectest} shows the power spectra measured from the
mock APM cluster catalogues, using the estimator described in Section
2. The crosses show the average value of $P(k)$ measured from $27$
mock surveys for each ensemble using three randomly chosen centres for
each simulation. The error bars show the standard deviation of the
mean $P(k)$, computed from the scatter over the $27$ mock surveys.
The circles in Figure~\ref{plotpsrectest} show the mean redshift-space
power spectrum for each ensemble computed from the full $N$-body
simulation volume ({\it i.e.} with no application of a selection
function or angular sky mask).
The values of $P_{w}(k)$ used in the weighting function were $P_{w}(k)
= 50000 \left(\hmpc\right)^{3}$ and $P_{w}(k) = 16000
\left(\hmpc\right)^{3}$ for the LCDM and SCDM simulations
respectively. These values of $P_{w}(k)$ were chosen to provide a 
match to the mean amplitude of $P(k)$ from these two cosmological models
over the range of wavenumbers plotted in the figure. The figure shows
that our estimator of $P(k)$ based on equation (1) 
provides a good measure of the power
spectrum of APM-like cluster catalogues in the LCDM model for
wavenumbers $k \gqt 0.02 \hmpcrev$. For the SCDM model, we see
some differences  in the power spectra estimates at
wavenumbers $0.02$ -- $0.05 \hmpcrev$, but these are not
large enough to eliminate the break in the power
spectrum estimates for the mock APM catalogues.

TE96 give a detailed analysis of biases in estimating  the power
spectrum from redshift surveys. There are two effects that can lead to
systematic differences between the estimated and true power spectra on
large scales: (a) the estimated power spectrum is a convolution of the
true power spectrum and the power spectrum of the window function of
the survey; (b) the estimated power spectrum is biased low at small
wavenumbers if the mean space density of the sample is estimated from
the survey itself. Both effects depend on the geometry of the survey
and on the form of the true power spectrum and will affect the
estimates of $P(k)$ from the mock APM catalogues plotted in
Figure~\ref{plotpsrectest}. However, these biases are not present in
the power spectrum estimates from the full periodic volumes of the
N-body simulations (plotted as the circles in
Figure~\ref{plotpsrectest}).  Thus, the close agreement between the
power spectra in each of Figures~\ref{plotpsrectest}(a) and
~\ref{plotpsrectest} (b) shows that neither bias is significant for
wavenumbers $k \gqt 0.02 \hmpcrev$.

The inset in the left hand panel of Figure~\ref{plotpsrectest} shows a
test of the error estimator used in this analysis. The crosses show
the error bars calculated from equation (2.4.6) of FKP, averaged over
$27$ realizations. The errors on each of these points show the error
on the mean. The triangles show the error on $P(k)$ calculated by
evaluating the spread in the measured value of $P(k)$ over the $27$
realizations of the APM cluster survey. Thus equation (2.4.6) of FKP
provides an excellent estimate of the true error on $P(k)$ for all but
the largest wavenumbers plotted in the figure.

\begin{figure*}
\centering
\begin{picture}(300,250)
\includegraphics{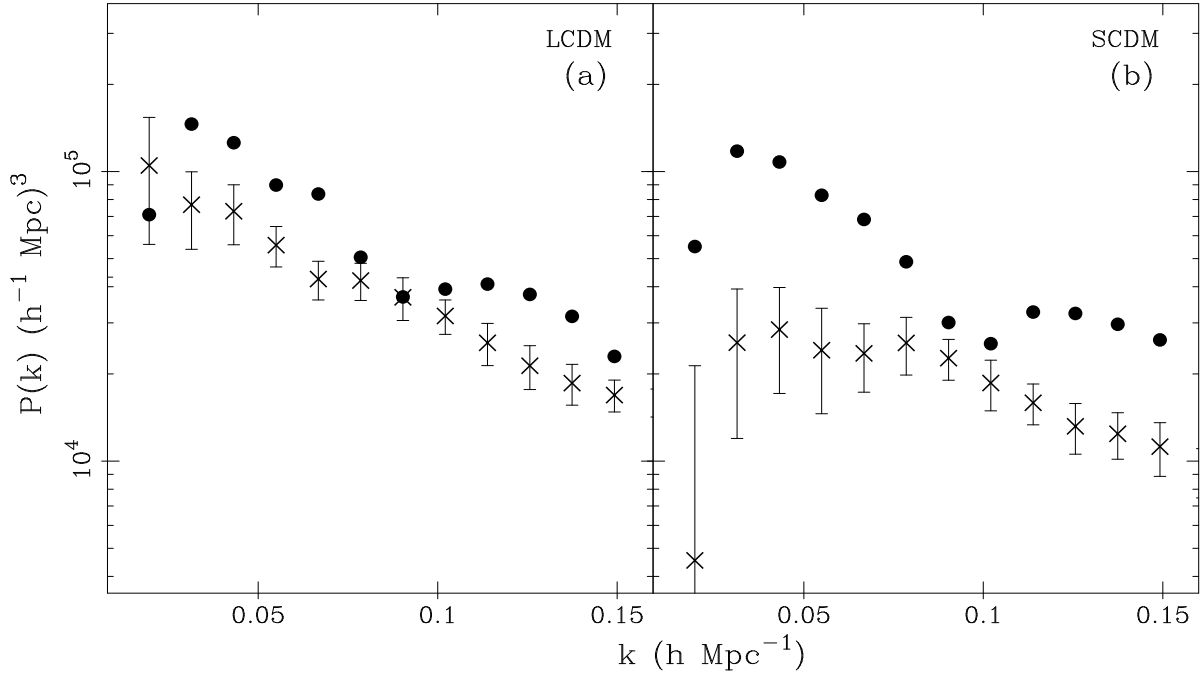}
\end{picture}
\caption[junk]{\label{compth} A comparison of the power spectrum of
APM clusters with the predictions of the LCDM and SCDM models. The
crosses in each case show the average power spectra of $27$ mock APM
cluster surveys extracted from the N-body simulations, together with
$1\sigma$ errors appropriate to a single mock survey. The circles show
the power spectra computed for the real APM cluster survey assuming a
background cosmology that is consistent with the theoretical models.}
\end{figure*}

In Figure~\ref{compth} we compare the power spectrum of the APM
clusters  with the power spectra measured from the mock
catalogues described above. The points for the real survey
differ slightly in the two panels because in each case we assume 
background cosmologies consistent with the LCDM and SCDM 
theoretical models ({\it cf}
Figure 2b). The APM cluster power spectrum
cannot be matched by the SCDM model, in agreement with our previous
analyses of the spatial two-point correlation function of the APM 
cluster survey (\pcite{DEMS92}, \pcite{DCESMD94}). The 
LCDM model is much closer to the observations in shape and amplitude, 
but under-predicts the clustering strength and does not reproduce
the turnover in the observed power spectrum at $k \simlt 0.03 \hmpcrev$.
A scale-invariant CDM  model with a lower value of 
$\Gamma$ ($\Gamma \approx 0.15$) would likely provide a better
match to the clustering amplitude of rich clusters at wavenumbers
$k \sim 0.04 \hmpcrev$ (e.g. see Figures $8$ and $9$ of \pcite{GCD95})  
but would fail even more strongly to match the downturn in the observed
power spectrum at small wavenumbers. 

These discrepancies between theory and observation do not seem
particularly problematic, given the uncertainties in both the
observational estimates and theoretical models. Neither the SCDM or
LCDM models considered here provide an acceptable match to the
observations, but these belong to a highly restricted class of CDM
model. It may be possible to achieve a better match with a CDM-type
model by allowing other parameters to vary, {\it e.g.} by
incorporating a tilt in the primordial spectral index
(\pcite{WSSD95}), or by invoking an admixture of hot and cold dark
matter (\pcite{KHPR93}).

\section{Discussion}

\begin{figure*}
\centering
\begin{picture}(300,280)
\includegraphics{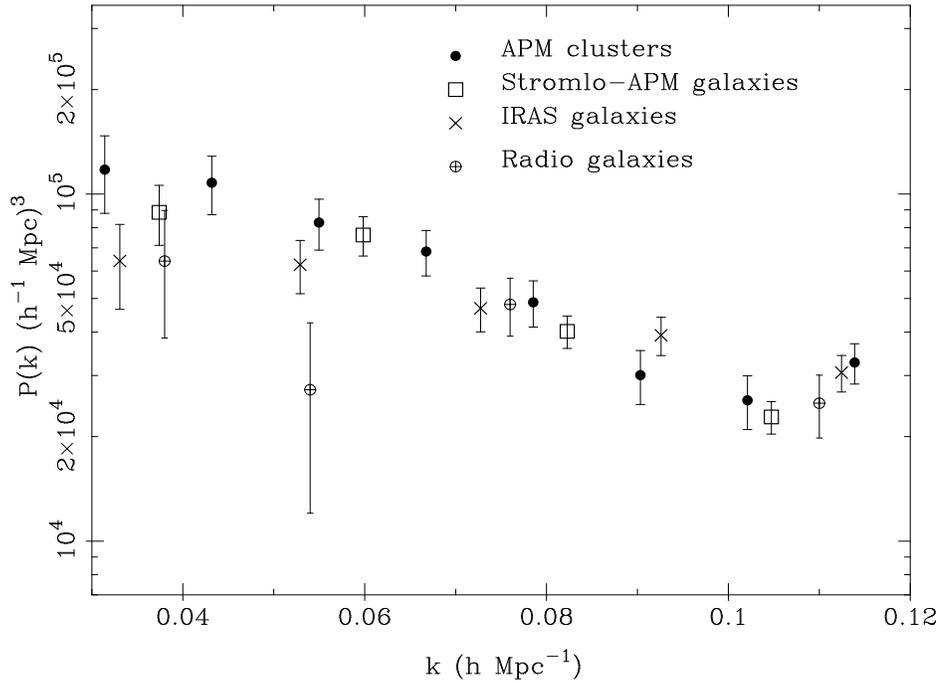}
\end{picture}
\caption[junk]{\label{compps} The power spectrum of APM clusters together with that of IRAS galaxies, optical galaxies and radio galaxies. The galaxy power spectra have been scaled upward by a scale-independent factor (see Table 1)
to match the amplitude of the APM cluster power spectrum.} 
\end{figure*}

\begin{table*}[t]
{\centerline{\bf Table I}}
\centering
\begin{tabular}{||c|c|c|c|c||}
\hline 
 & $k$ $\hmpcrev$  & $b_c^{2}$ & $\chi^{2}/d.o.f$ & P  \\ \hline \hline
IRAS & 0.05 - 0.1 & 6.43 $\pm$ 0.73 & 1.65 & 0.18 \\ 
Stromlo-APM & 0.05 - 0.1 & 3.13 $\pm$ 0.32 & 0.19 & 0.90 \\ 
Radio gals & 0.04 - 0.1 & 0.99 $\pm$ 0.15 & 1.34 & 0.25 \\ \hline
\end{tabular}    
\caption{\label{bias} Values for the relative bias (squared) between
IRAS, optical and radio galaxies compared to APM clusters. Column 2
gives the range in wavenumber over which the fit was performed. Column
3 gives the factor by which the galaxy power spectrum must be scaled
to match the clusters power spectrum i.e. the relative bias squared.
The last two columns give the $\chi^{2}$ per degree of freedom and
the probability for the fit to a scale-independent relative bias.}
\end{table*}

We first compare the power spectrum of rich clusters with those for
other types of object in the Universe.  Figure~\ref{compps} shows the
APM cluster power spectrum together with power spectra for 
optical Stromlo-APM galaxies (\pcite{Lov92}, TE96), IRAS galaxies
(from the combined QDOT and $1.2Jy$ surveys, as calculated in
\pcite{tad95}) and a sample of $310$ radio galaxies (as calculated by
\pcite{PN91}). The spectra for optical, radio and IRAS galaxies have
been scaled upwards by constant factors $b_c^2$ to match the
amplitude of the APM clusters power spectrum. The factors $b_c$ are thus
the relative bias parameters for the three samples and are computed by
performing a weighted fit to the ratio of the observed power spectra
over a specified range in wavenumber.  Table~\ref{bias} lists the
range in wavenumber over which the data were fitted, the values of the
relative bias factors $b_c$, as well as the probability for the fit. A
reasonable value for the fit probability shows that these power
spectra are consistent with the hypothesis of a linear relative bias
between the various tracer objects over the scales indicated in Table
1.

\scite{PD94} found a relative bias parameter of $2.37$
between radio galaxies and Abell clusters.  This differs from the
lower relative bias between radio galaxies and APM clusters listed in
Table 1.  This is a result of the higher amplitude of the power
spectrum of Abell  compared to APM clusters.  However, there
is now a large body of evidence to show that the clustering of Abell
clusters is enhanced by non-uniformities in the Abell catalogue and
does not reflect the true clustering properties of rich clusters
(\pcite{S88},
\pcite{EDSM92}, \pcite{Richclus}). The relative bias of $b_c = 0.99$ 
between radio galaxies and APM clusters from Table~\ref{bias} is thus 
consistent with the idea that luminous
radio galaxies preferentially inhabit the cores
of rich clusters of galaxies.

\begin{figure}
\centering
\begin{picture}(180,200)
\includegraphics{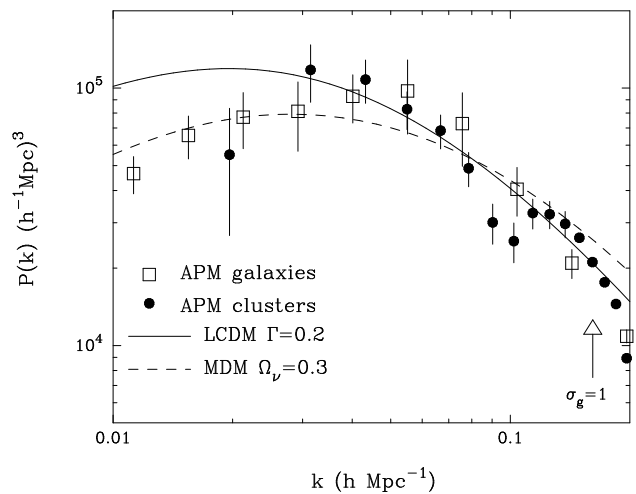}
\end{picture}
\caption{\label{gpe} The filled circles show the power spectrum of APM
clusters as plotted in Figure 6. The open squares show the power
spectrum of APM galaxies estimated by inverting the two-dimension
angular correlation function as described by Maddox et al. 1996 (see
their Figures 32 -- 35). The APM galaxy power spectrum has been
multiplied by a factor of $5$ to match the amplitude of the cluster
power spectrum. The error bars were determined from the scatter in the
inversions from four nearly equal area zones of the APM survey. The
two lines show linear theory power spectra for a scale-invariant
LCDM model with $\Omega_\Lambda = 0.8$, $h=1$ (solid line) and a
mixed dark matter model with $\Omega_\nu = 0.3$, $h=0.5$.  The linear
spectra have been normalised arbitrarily to match the cluster power
spectrum approximately in the region $k \sim 0.05 \hmpcrev$.}
\end{figure}

The agreement between the shapes of the power spectra for rich
clusters of galaxies, IRAS and optically selected galaxies is
consistent with a simple linear bias model. This provides an argument
against models in which the galaxy formation process introduces
spatial correlations so that galaxies are non-linearly biased with
respect to the mass distribution (see {\it e.g.}
\pcite{Bab91}). Figure~\ref{compps} suggests a simpler interpretation,
in which different types of galaxies and rich clusters are linearly
biased on large-scales and each traces the shape of the underlying
matter power spectrum.

We have found some evidence from the power spectrum of the APM cluster
sample for a peak at $k \sim 0.03 \hmpcrev$, followed by a downturn at
smaller wavenumbers. The tests described in Section 3 suggest that the
biases in the estimator of $P(k)$ are too small to cause such a
downturn. Figure 7 shows the APM cluster power spectrum (filled
circles) together with the three-dimensional power spectrum determined
by inverting the angular two-point correlation function measured for the APM
Galaxy Survey (open squares). The latter points are from the analysis
of Maddox \etal (1996) who applied the inversion technique developed
by Baugh and Efstathiou (1993). The amplitude of the APM galaxy power
spectrum has been multiplied by a factor of $5$ to match the amplitude
of the cluster power spectrum. Both the cluster and galaxy power
spectra show a peak at a wavenumber $\sim 0.03$--$0.04
\hmpcrev$, which suggests that the peaks reflect a real feature
of the underlying mass distribution. As mentioned in the introduction,
such a peak must exist at a  wavenumber $\simgt 0.01 \hmpcrev$
since the anisotropies of the cosmic microwave background radiation
on angular scales $\simgt 1^\circ$ are well approximated by a 
scale-invariant initial spectrum
(\pcite{Bond96}). 

The lines in Figure 7 show linear theory power spectra for two
models, normalized to match the observations at a wavenumber $k \sim
0.05 \hmpcrev$. The $\Gamma=0.2$ LCDM model has too much power on
large scales to provide a good match to either the cluster or galaxy
power spectra. The dashed line shows a scale-invariant mixed dark
matter (MDM) power spectrum, with parameters $\Omega_\nu=0.3$,
$\Omega_b = 0.05$, $\Omega_m = 0.65$, and $h=0.5$ (where $\Omega_\nu$,
$\Omega_b$ and $\Omega_m$ are the respective cosmological densities in
massive neutrinos, baryons and cold dark matter). The power spectrum
of the MDM model is computed from the fits given in \scite{Ma96}, but
over the range of wavenumbers plotted in Figure 7 there is little
difference between the MDM spectrum and a CDM spectrum of the form (4)
with $\Gamma = \Omega_m h
\sim 0.3$. As Figure 7 shows, a CDM-like model with $\Gamma \sim 0.3$
has a peak in the power spectrum that approximately matches the
observations. 

Because of the low space density of rich clusters of galaxies and
their enhanced clustering strength compared to normal galaxies, rich
clusters provide a powerful probe of large-scale structure in the
Universe. At wavenumbers $k \simgt 0.01 \hmpcrev$, redshift surveys of
only a few thousand rich clusters are capable of providing comparable
results on the power spectrum, to those from redshift surveys of order
$10^6$ galaxies.  Figure 7 thus suggests that a modest increase in the
size of the cluster redshift sample could confirm the reality of a
peak in the power spectrum at a high significance level.

{\bf Acknowledgments:} We thank John Peacock for communicating the
power spectrum of radio galaxies in a machine readable form.

\end{document}